\documentclass[letter,scriptaddress,twocolumn, showkeys]{revtex4}
\usepackage{ amssymb }
	\usepackage{amsmath}
	\usepackage{makeidx}
	\usepackage{amsfonts}
	\usepackage[ansinew]{inputenc}
	\usepackage[usenames,dvipsnames]{pstricks}
	\usepackage{subfigure}
	\usepackage{epsfig}
	\usepackage{pst-grad} 
	\usepackage{pst-plot} 
		\usepackage{amsthm}
\usepackage{ amssymb }
 	\usepackage{makeidx}
	\usepackage{amsfonts}
 	\usepackage{listings}
 	\usepackage{mathtools}
	\usepackage{float}
	
	\usepackage[colorlinks,hyperindex]{hyperref}
	\hypersetup
	{
		colorlinks,%
		citecolor=blue,%
		linkcolor=blue,%
		urlcolor=blue,%
	}
	\usepackage{tikz}

\usepackage{amsmath}
\usepackage{amssymb}
\usepackage{graphicx}
\usepackage{xcolor}
\usepackage{float,graphicx}
\usepackage{placeins}
\usepackage{color, colortbl}
\usepackage[colorinlistoftodos]{todonotes}

\theoremstyle{definition}

	\theoremstyle{plain}

\makeindex

\begin{document}

\title{Extrapolating continuous color emotions through deep learning}

\author{ Vishaal Ram$^{a}$, Laura P. Schaposnik$^{\star, b}$, 
Nikos Konstantinou$^{c}$, 
Eliz Volkan$^{d}$,
Marietta Papadatou-Pastou$^{e}$,  
Banu Manav$^{f}$,   Domicele Jonauskaite$^{g}$,   Christine Mohr$^{g}$
}
  \affiliation{($\star$) Corresponding author: schapos@uic.edu}

\begin{abstract}
By means of  an experimental dataset, we use deep learning to implement an RGB  extrapolation of emotions associated to color, and do a mathematical study of the results obtained through this neural network.  In particular, we see that  males (type $m$ individuals) typically associate a given emotion with darker colors while females (type $f$ individuals) with brighter colors. A similar trend was observed with older people and associations to lighter colors. Moreover, through our classification matrix,  we  identify which colors have  weak associations to emotions and which colors are typically confused with other colors. 
 \end{abstract}

\maketitle
 
\section{Introduction}
The relation between colours and human emotion has been studied for more than a century (e.g., see for instance \cite{past1,past2,past3,mas1,mas2,mas3,mas4, mas7}). Even longer ago, colours were commonly associated to emotions in a universal manner that allowed populations to understand quickly the given emotions.  For example, for centuries in many cultures  it has been said that someone ``had the blues" \footnote{See, for instance  John James Audubon ("The Birds of America")  journal from 1827 where he wrote that he "had the blues".} or ``is feeling blue" when being down or sad. As explained in \cite{website}, the phrase ``feeling blue" comes from deepwater sailing ships:  If a ship lost the captain or any of the officers during its voyage, then blue flags would be shown, and   a blue band  would be painted along the entire hull when returning to home port.

Inspired by \cite{jonauskaite2019sun,mas8} we consider their data base \cite{plu1} to analize the correlation between colours and emotions via a deep learning approach. Whilst machine learning techniques have been used before in this direction (e.g. see \cite{plu2} and references therein), we take a novel approach which allows us to discern several interesting patters.

 When using a deep learning approach to quantify color-emotion associations, one expects to observe certain behaviors. In particular:
\begin{itemize}
    
    \item[(a)] Certain colors should have strong associations with emotion and a high classification accuracy. 
    \item[(b)] Some colors   are associated with multiple emotions would have a low classification accuracy. 
    \item[(c)] Regional and geographic factors may have a factor in color emotion associations and would provide a deep learning approach to distinguish region.  
    \item[(d)]  Several colors are associated with a single emotion. 
    \end{itemize}
    
Colour association studies usually consider a discrete number of colours. In particular, this is the case of the study leading to the dataset \cite{plu1} which we shall use in the present paper, where participants associated emotions to 12 colour terms: red, orange, yellow, green, blue, turquoise, purple, pink,
brown, black, grey, and white. It should be emphasised that the experiment did not show colours but rather gave the terms of colours and it was left to the participants imagination the choice of what those words meant. 

\begin{figure}[H]
  \centering \includegraphics[scale = 0.06]{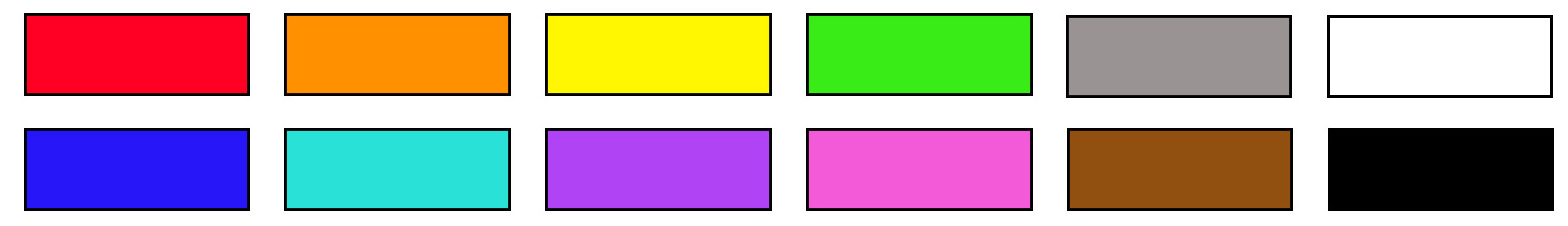}
  \caption{A depiction of the 12 colors used in \cite{plu1}. }
      \label{col1}
\end{figure}

To carry out our mathematical study, we have used the standard Decimal Code (R,G,B) to represent the 12 colours of \cite{plu1}, a depiction of which is in Figure \ref{col1}.

\begin{figure}[H]
  \centering \includegraphics[scale = 0.2]{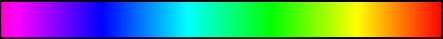}
  \caption{A depiction of the  continuous RGB palate used in the present paper. }
      \label{col2}
\end{figure}

In the last decades colours have also been studied in terms of emotional reactions to color hue, saturation, and brightness (e.g.,   \cite{hue,mas6}). Here, we shall put the two approaches together to consider a novel path, where we let the colour association within our neural network take a continuum  of colours, hence considering a continuous RGB analysis \footnote{Where RGB stands for when red, green, and blue lights are added together in various ways to reproduce a broad array of colors.},   depicted in Figure \ref{col2}.

  After introducing some background in Section \ref{machine}, we dedicate Section \ref{results} and Section \ref{conti} to the main findings of our work, which  can be seen in two different directions in terms of associations of individuals of type $m$ (males) and of type $f$ (females):
\begin{itemize}
\item From the classification matrix: we used a neural network approach which is different from the paper's SVM approach \cite{plu2}:
\begin{itemize}
\item  From the matrix we can identify which colors have weak associations to emotions and which colors are typically confused with other colors. 
\end{itemize}

\item Through an RGB regression we study single emotion associations as well as how associations varied across age and gender ($m$ and $f$),  which we have not seen previously discussed through mathematical models.  In particular, we see that:

\begin{itemize}
\item Males typically associated the same emotions with darker colors than females. 
\item Older people do associations to lighter colors. 
\end{itemize}
\end{itemize}
We shall expand on the analysis and applications of the above findings in Section \ref{last}. 
\pagebreak
\section{Background: A machine learning approach to quantify the specificity of color-emotion}\label{machine}

The current research was inspired by the previous work \cite{plu2}, and thus we shall dedicate this section to reviewing some of the main results which shall prove useful for our research and for comparison with our results.  Their research proposes that color-specificity of emotion associations and country-specificity of color-emotion associations can be measured using a multivariate pattern classification approach. 
       When classifying the data,  the authors used an optimized SVM (support vector machine) approach with a 10-fold cross-validation (CV) to evaluate accuracy and considered several classifiers. 
             \begin{itemize}
        \item
        The first classifier predicted color on the basis of 20 ratings of color-emotion associations. The classifier achieved an accuracy of 38.7\% when tested on 4 countries, and achieved an accuracy of 30.4 \% when the classifier was applied to a data set of 30 countries. 
       True positive rate was the highest for black and red, followed by brown, pink, and grey. Thus they elicited very specific associations. 
        
        \item The second classifier predicted country on the basis of 240 ratings of color-emotion associations to quantify the degree to which color-emotion associations are country specific. The same learning model was used and achieved an accuracy of 80.2 \% when tested on the 4 countries China, Greece, Germany, and UK. 
         The data indicated strong country-specific color-emotion associations. (e.g. association between brown and disgust was stronger in Germany than the other remaining countries, and almost non-existent in China, and participants in Greece were the only to indicate a strong association between purple and sadness).
    \end{itemize}
    
\begin{figure}[H]
  \centering \includegraphics[scale = 0.3]{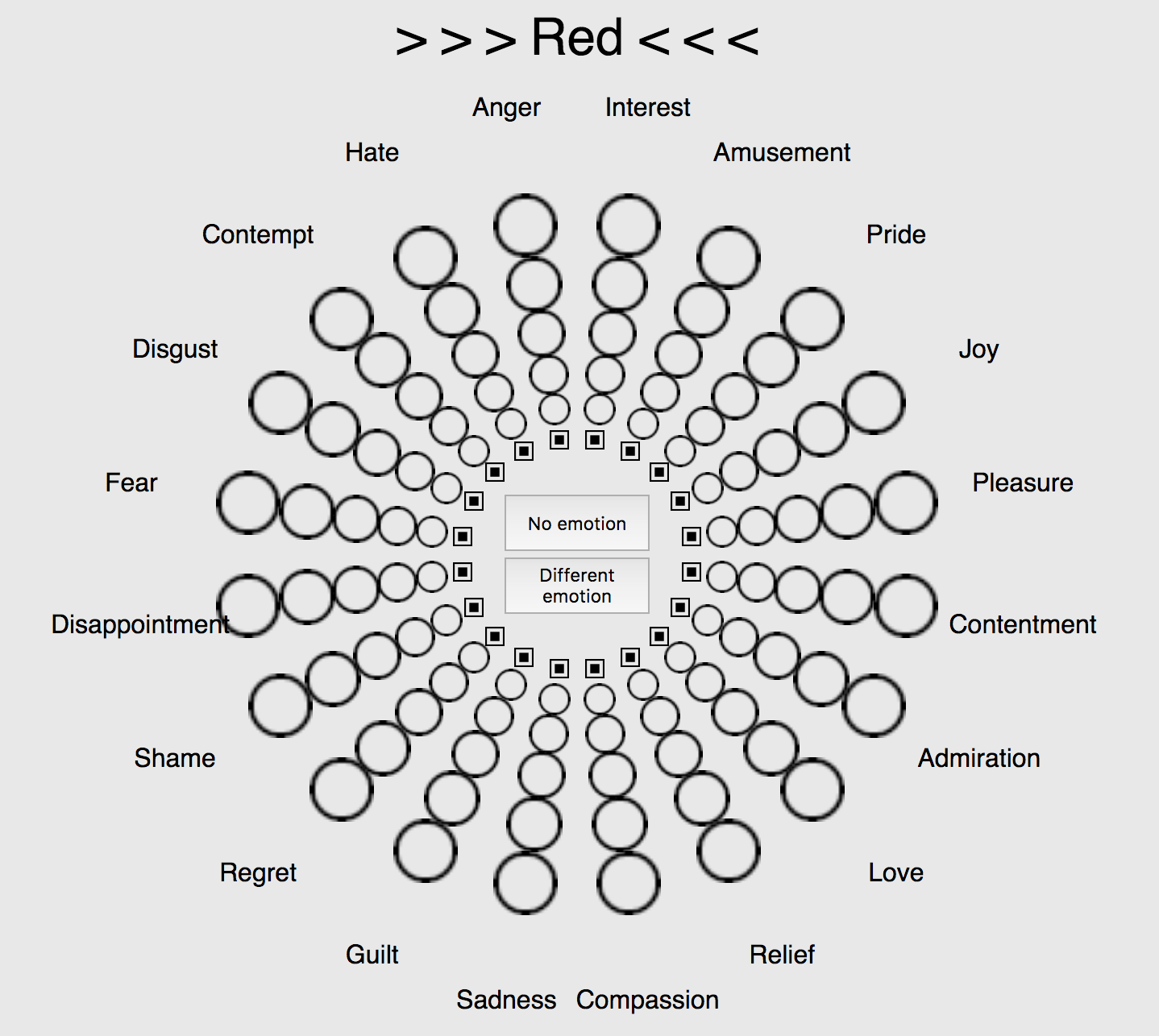}
  \caption{Example of the platform within the survey used to collect the data in \cite{plu1}. }
      \label{fig:disc}
\end{figure}

       The data used in the project comes from Forsbase's (2019)  dataset \cite{plu1} obtained through an International colour-emotion survey program conducted by   Domicele Jonauskaite and  Christine Mohr (PI). 
 Since 2015, researchers from the Institute of Psychology at the University of Lausanne have been collecting data from an international color-emotion survey.  The data is collected online via the online platform  \href{https://www2.unil.ch/onlinepsylab/colour/main.php}{https://www2.unil.ch/onlinepsylab/colour/main.php}, through which participants indicated the emotion they associated to different colour words through a graph of options, as shown in Figure \ref{fig:disc}.

  The survey asks for basic information including age, gender, country of origin and language. Then the user is presented with 12 colour terms and is asked to choose one, several, or none of the 20 emotions they associate with these colour terms. Participants also rate the intensity of each associated emotion, quantified by an integer from 0 to 5. In the present paper, we shall use the data for  a sample of Greek speakers from Greece, Greek speakers from Cyprus, Turkish speakers from Cyprus, and Turkish speakers from Turkey comprising 944 participants.             

%

    \section{Predicting the associated colours}\label{results}
     
\begin{figure}[H]
  \centering \includegraphics[scale = 0.25]{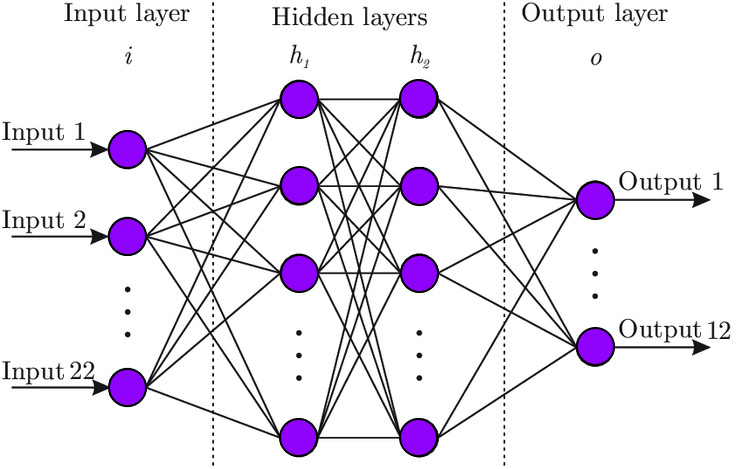}
  \caption{Diagram of the neural network used.}
      \label{fig:net}
\end{figure}

      As mentioned before, the data we shall use in the present manuscript  consists of 944 submissions from \cite{plu1}. We shall consider the gender variables as indicating participants' type $m$ (male) and $f$ (female), and shall consider the standard  Decimal Code (R,G,B) conversion of the colours' words used in \cite{plu1} to carry out our mathematical study.
      Including age and gender, a total of 22 input variables have been used to predict the associated color. The learning structure used was a neural network with two hidden layers of size 10, and an  output layer with 12 nodes, one for each color as depicted in Figure \ref{fig:net}.

\begin{figure}[H]
    \begin{center}
    \includegraphics[scale = 0.7]{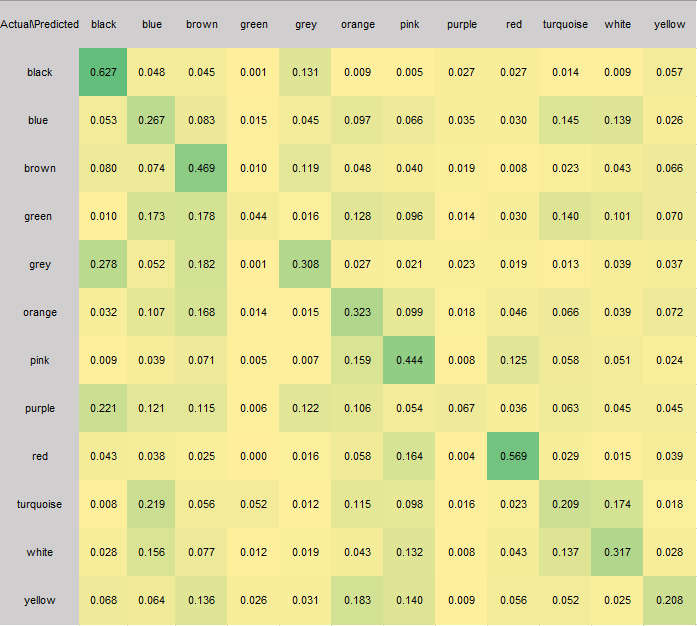}
    \end{center}
    \caption{Confusion Matrix.
    }
    \label{fig:conf2}
    \end{figure}

      The predicted color is the output node with the greatest activation. 
 The neural network was trained using a training set of size 2000 and testing using a cross-validations set of size approximately 10000. We used the standard back-propagation algorithm to optimize the edge weight parameters of the neural network to improve the classification accuracy on the cross-validation set. Their results are summarized in  Figure \ref{fig:conf2}, with classification accuracy at about 33 \%. 
\subsection{Accuracy depending on colours}
\begin{figure}[H]
  \centering \includegraphics[scale = 0.22]{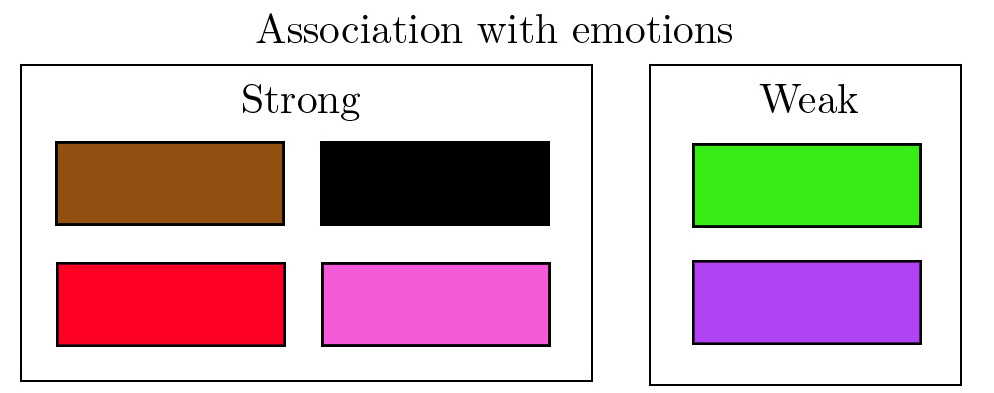}
  \caption{Emotional association within discrete colour analysis}
      \label{assoc}
\end{figure}

    From our study we see different levels of accuracy obtained in our predictions, allowing us to infer the level of emotional association that colours have. Indeed, the following is observed, as depicted in Figure \ref{assoc}:
    \begin{itemize}
    \item The colours black, brown, pink, and red are predicted with high accuracy and indicate a strong association with emotion;
    \item In contrast, the colors green and purple are predicted with very low accuracy and indicate weak association with emotion. 
   \end{itemize} 

   \subsection{Colour exchanges}
\begin{figure}[H]
  \centering \includegraphics[scale = 0.13]{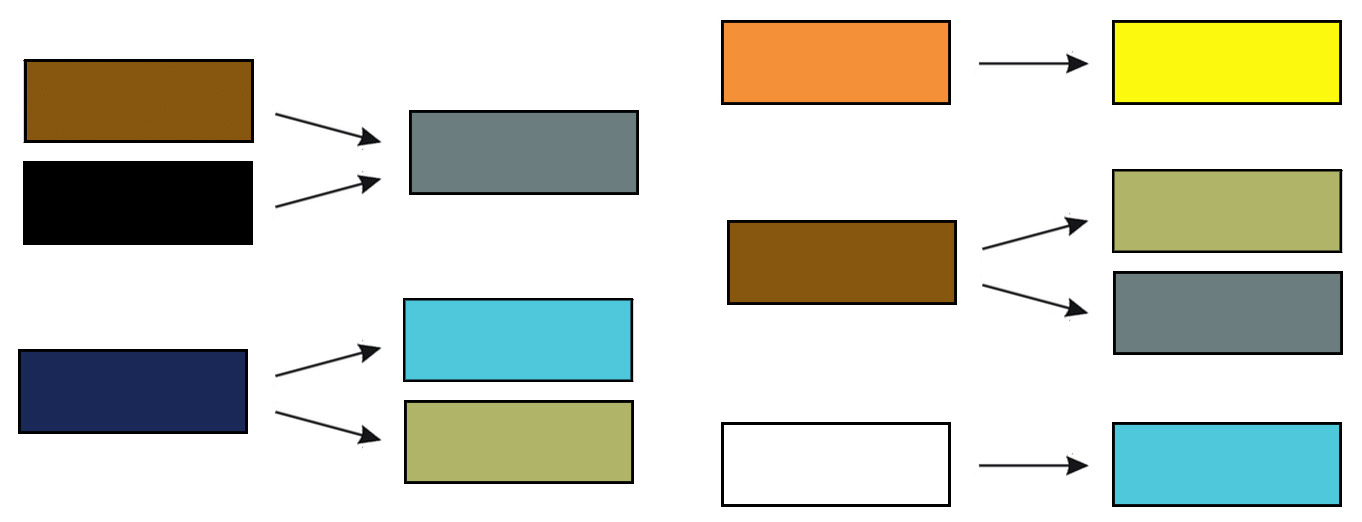}
  \caption{Confusion exchanges}
      \label{flip}
\end{figure}

Through our study we note that some pairs of colors are frequently confused, indicating the variability of the emotional association, and the need for further understanding the relation between colours in such pairs both from the visual as well as emotional point of view. In particular we note that, as depicted in Figure \ref{flip}, the following occurs:
   \begin{itemize}
   \item black and brown are confused for grey;
   \item  blue for turquoise and green;
   \item orange for yellow;
   \item  brown for green and grey;
   \item and white for turquoise. 
    \end{itemize}
    
 Interestingly, colors like green and purple are rarely predicted at all, even though all colors appear the same number of times. This suggests that their association to emotion is weak enough to make it beneficial for the neural network to always output a different color. 
      In order to improve accuracy, and given the above usual exchanges of colours, we consider the following two alternative set ups to understand the colour association:
    \begin{itemize}
        \item[(A)] Group similar colors  to minimize confusion and improve accuracy;
        
        \item[(B)]  Implement a continuous RGB output instead of a classification network. This can be used to demonstrate how subtle changes in color associations are influenced by varying emotion, and it should be noted that it was not attempted in \cite{plu2}.
     \end{itemize}


\subsection{Grouping Colors}
\begin{figure}[H]
    \begin{center}
    \includegraphics[scale = 0.6]{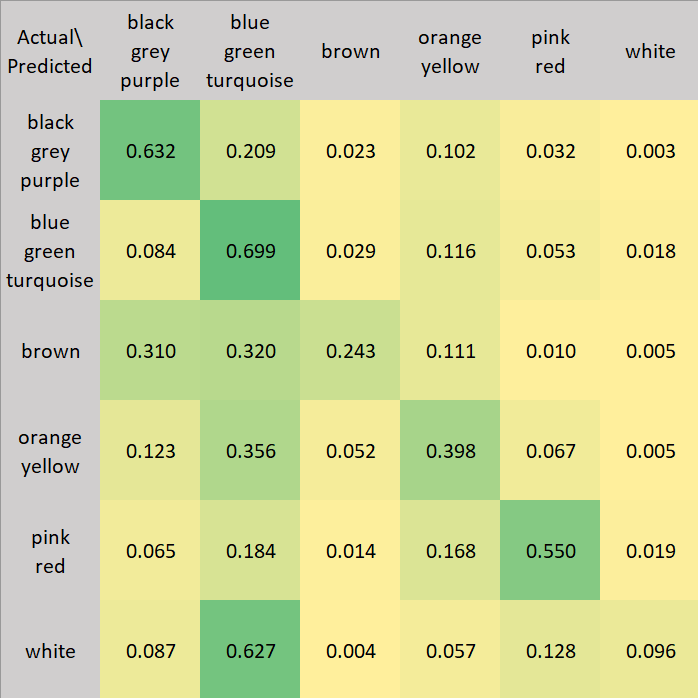}
    \end{center}
    \caption{Confusion Matrix for Color Groupings.
    }
    \label{fig:conf1red}
    \end{figure}

 To improve the consistency of the color classification, one natural step is to reduce the dimension of the output layer by grouping colors by similar emotion associations in accordance to Figure \ref{fig:conf1red}. In particular, we have grouped together the darker colors black, grey, and purple. By reducing the dimension by a factor of two, we have significantly improved the classification accuracy to $52 \%$ from the $33 \%$ in the full color classification. One notable observation is that the brown group is more often classified as the black-grey-purple group and the blue-green-turquoise group than its own group, a pattern not observed in the classification without dimension reduction. Additionally, the white group is almost never classified correctly, most often being classified as the blue-green-turquoise group. One possible explanation being that white is mostly associated with relief, a trait that is shared among the blue-green-turquoise group. Lastly, the orange-yellow group was incorrectly classified as the blue-green-turquoise group about just as often as it was correctly classified, which again can be explained by shared emotion such as joy and pleasure.

  \begin{figure}[H]
    \begin{center}
    \includegraphics[scale = 0.9]{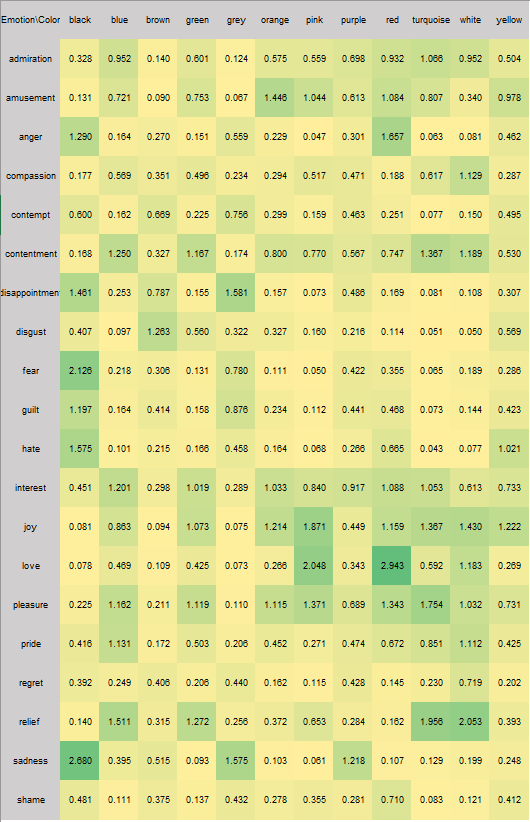}
    \end{center}
    \caption{Average Color-Emotion Association}
    \label{fig:cetable}
    \end{figure}
 \subsection{Average colour association}
        
   In the survey, participants are asked to describe the strength of the association between color-emotion pairs on a scale from 0 to 5. 
      The table shown in Figure \ref{fig:cetable} displays the average association value across every color-emotion pair among all the participants in the data set, which serves as a visualization of the entire data set without the age and gender factors. Some of the limitations of the results described in the single color association is that certain emotions can strongly associate with multiple conflicting colors, leading to an ambiguous mix of colors, often producing a brown. For example, both amusement and pleasure are associated with both colors like blue and turquoise as well as colors like red and orange,   the combination of which cannot be described by a single color, leading to some ambiguity and variation. 

Certain colors such as black, pink, and red display strong associations with at least one emotion, while others, such as brown and yellow, display weaker associations across all emotions. Likewise, we see that certain emotions such as love, relief, and sadness display strong associations with at least one color, while others, such as contempt, regret, and shame, display weaker associations across all colors.\\

    \section{Continuous RGB analysis}\label{conti}
    
    The use of a continuum of colours for studying human emotional associations has appeared in the literature for many decades now  (see, for example see the use within neural networks in \cite{color3}, within emotional expressions in robotics \cite{color2}, and when considering color combinations in \cite{color1}),  and has become more important recently because of their impact in image retrieval processes.  Whilst many experiments have used   color emotion metrics  for single colors or pairs of colors, in many cases similar   metrics were recently used in image retrieval showing that  humans perceive color emotions for multi-colored images in similar ways (e.g. see \cite{solli2011color}). Hence, it becomes relevant to analize the continuum of colours in the context of \cite{plu1} to deduce novel correlations between colors and emotions.

\begin{figure}[H]
    \begin{center}
    \includegraphics[scale = 0.5]{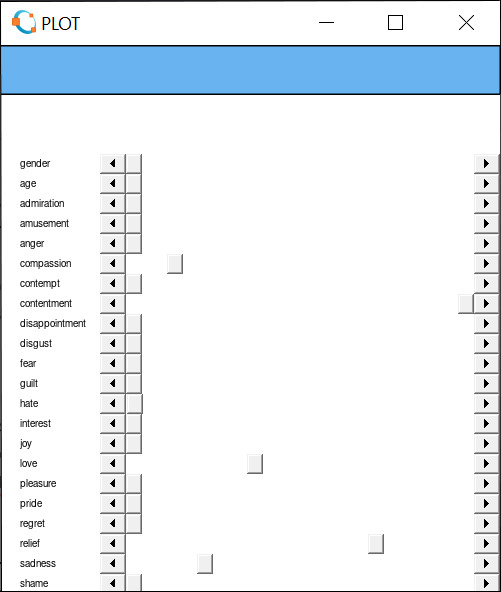}
    \end{center}
    \caption{RGB Regression Interface}
    \label{fig:RGB}
    \end{figure}

      The RGB regression study was implemented with the same neural network depicted in Figure \ref{fig:net}, with 22 nodes, two hidden layers of size 20 and 10, and 3 output nodes, representing the intensity of red, green, and blue. The training set (size 2000) was modified so that each color was converted to an RGB value based on standard conventions for color.
       By means of the neural network described above, we develop an interface which allows one to predict the colour depending on the choice of different variables. This interface, shown in Figure \ref{fig:RGB}, has a slider   for each input variable which can be adjusted. The age slider varies from 0 to 50, the gender slider varies from 0 to 1 (0 being female while 1 male), and each emotion varies from 0 to 5.     The color at the top represents the predicted color based on the RGB output. 

\subsection{Color-emotion associations across gender}

Gender differences when encoding and decoding color associations and facial emotions have long been considered (e.g. see \cite{rotter1988sex} and references therein). Whilst in many cases differences have been found, in other settings gender and age have not presented much difference. For example, gender and age were shown to be determining factors in the selection of achromatic black \cite{manav2007color}, whilst here were no sex differences in the main emotions linked to red within the children considered in \cite{boyatzis1994children}.

\begin{figure}
  \centering \includegraphics[scale = 0.45]{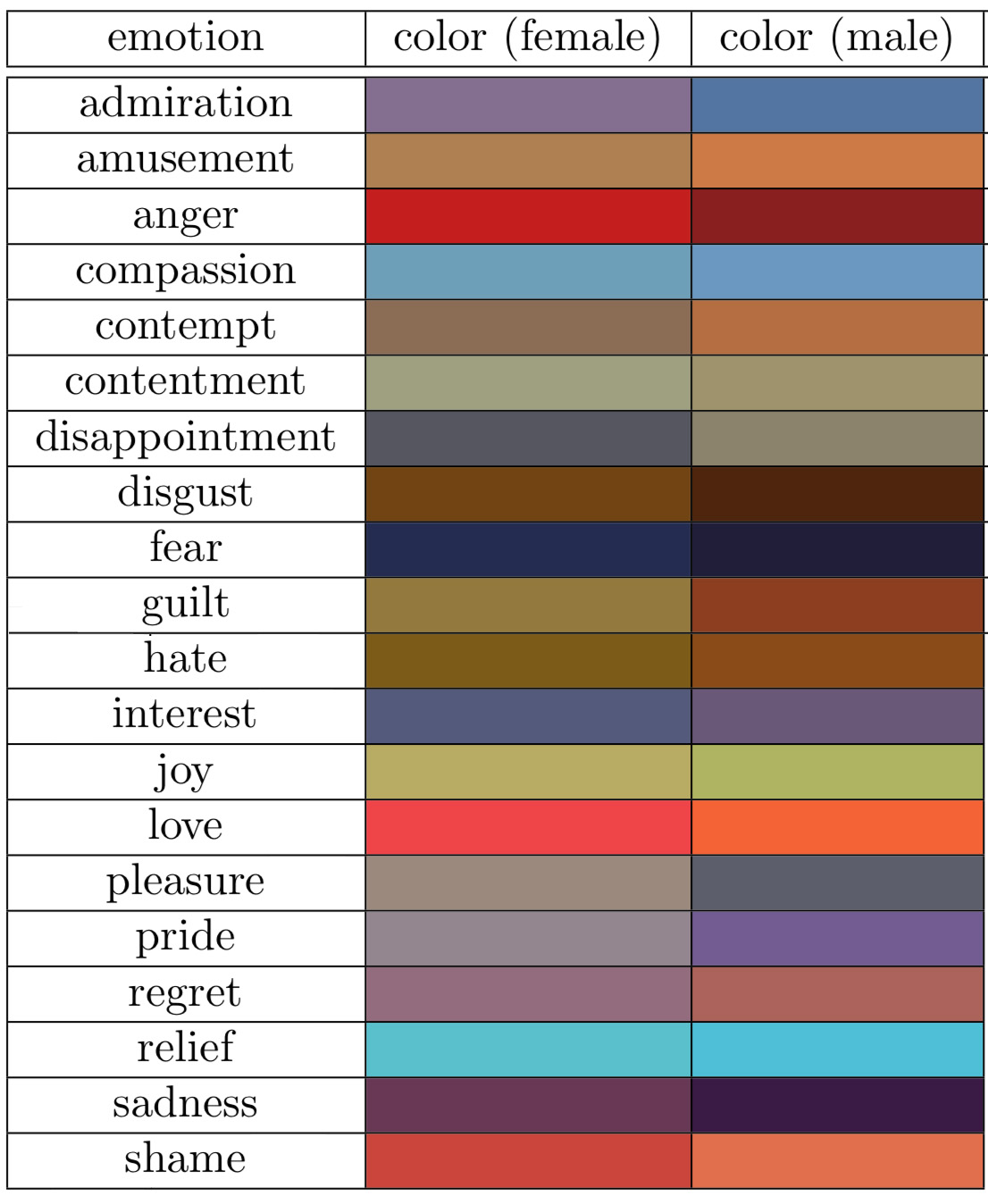}
  \caption{Gender (type $m$ and $f$) mathematical comparison in single color associations (for Age = 20)}
      \label{fig:table2}
\end{figure}

       
            By adjusting the gender input in the interface of Figure  \ref{fig:RGB}, one can compare color-emotion associations across gender as shown in Figure \ref{fig:table2}. Indeed, through the regression, we find that females associate admiration with a dark purple, while males a dark blue. Across most emotions, we find: 
            \begin{itemize}
            \item both genders typically associate with similar  emotions and colors.
            \item exceptions include admiration, in which females associate with a light purple while males with a blue, 
            \item and regret, which females associate with a dark pink while males with a red-orange. 
            \end{itemize}

Through our study we also found  that females typically associate emotions with lighter shades. This is most clearly seen in:   
\begin{itemize}
\item  anger, compassion, disgust, guilt, pleasure, pride, relief, and sadness;
\item   exceptions, in which cases males choose lighter colors, include shame and disappointment.
\end{itemize}
 Interestingly, this agrees with experiments done for young ages, where young boys were more likely to associate positive emotions with darker colours than girls, e.g.~see \cite{boyatzis1994children}.

\subsection{Color-emotion associations across age}

\begin{figure}[H]
    \begin{center}
    \includegraphics[scale = 0.48]{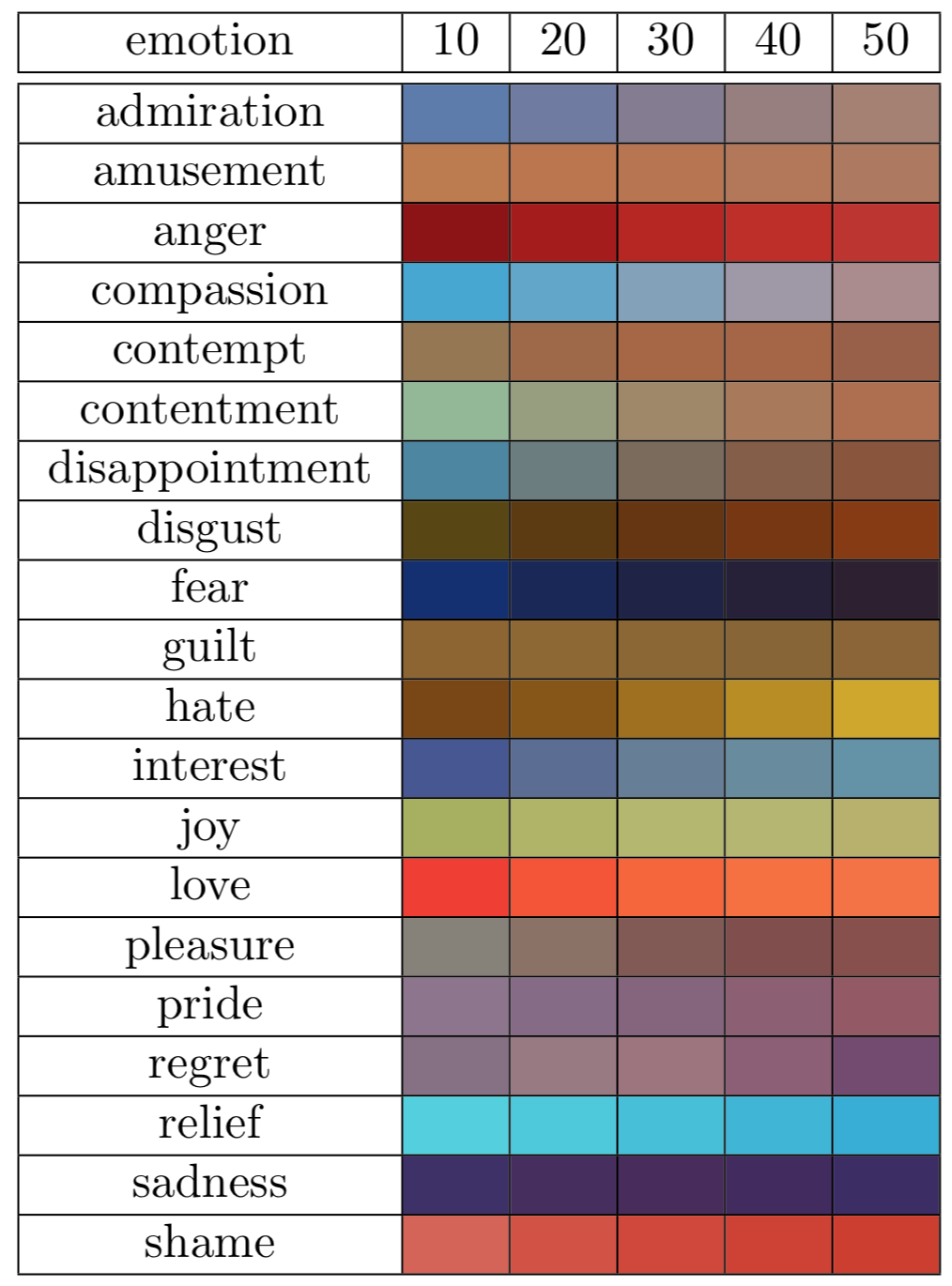}
    \end{center}
    \caption{Age comparison in single color associations (Gender Neutral), where shades of colours are obtained through the standard Decimal Code (R,G,B) obtained through our algorithm associated to each emotion. }
    \label{fig:table3}
    \end{figure}

  By adjusting the age input, we can compare color-emotion associations across age as shown in Figure \ref{fig:table3}. Through our mathematical model, we find that:
  \begin{itemize}
  \item  younger people associate anger with a darker shade of red, while older people a lighter shade;
  \item    older people associate  most emotions with lighter shades than younger people. Examples are   anger, compassion, hate, interest, and love,
  \item  on the other hand, older people associate relief with a darker blue.
  \item There exist  some emotions for which we see a consistent color across the age spectrum such as contempt, guilt, joy, sadness, and shame.
  \end{itemize}
  
    Some emotions shifts are associated with a more brown color as age increases, as in admiration and contentment, which could indicate ambiguity as in multiple color associations, or simply a lack of data as the age increases leading to extrapolation.   We shall return to this in Section \ref{last}, where we shall compare our results to those from experimental papers which have looked into aging and color-emotional associations \cite{phillips2002age,ebner2009young,zentner2001preferences}.

   It is interesting to note that in the present manuscript we have considered responses from Greek speakers from Greece, Greek speakers from Cyprus, Turkish speakers from Cyprus, and Turkish speakers from Turkey, and thus generically we get anger associated with shades of red. However, this should not hastily be generalized to a broader population: indeed, there have been studies such as \cite{hupka1997colors} showing that people from Poland connected anger, envy, and jealousy with purple instead of red. In their paper it is suggested that {\it ``cross-modal associations originate in universal human experiences and in culture-specific variables, such as language, mythology, and literature''}, and it would be indeed interesting to design a neural network to discern those differences once data from different regions is readily available.

  \pagebreak
\subsection{Emotions associated to single colors.}  
\begin{figure}[H]
    \begin{center}
    \includegraphics[scale = 0.36]{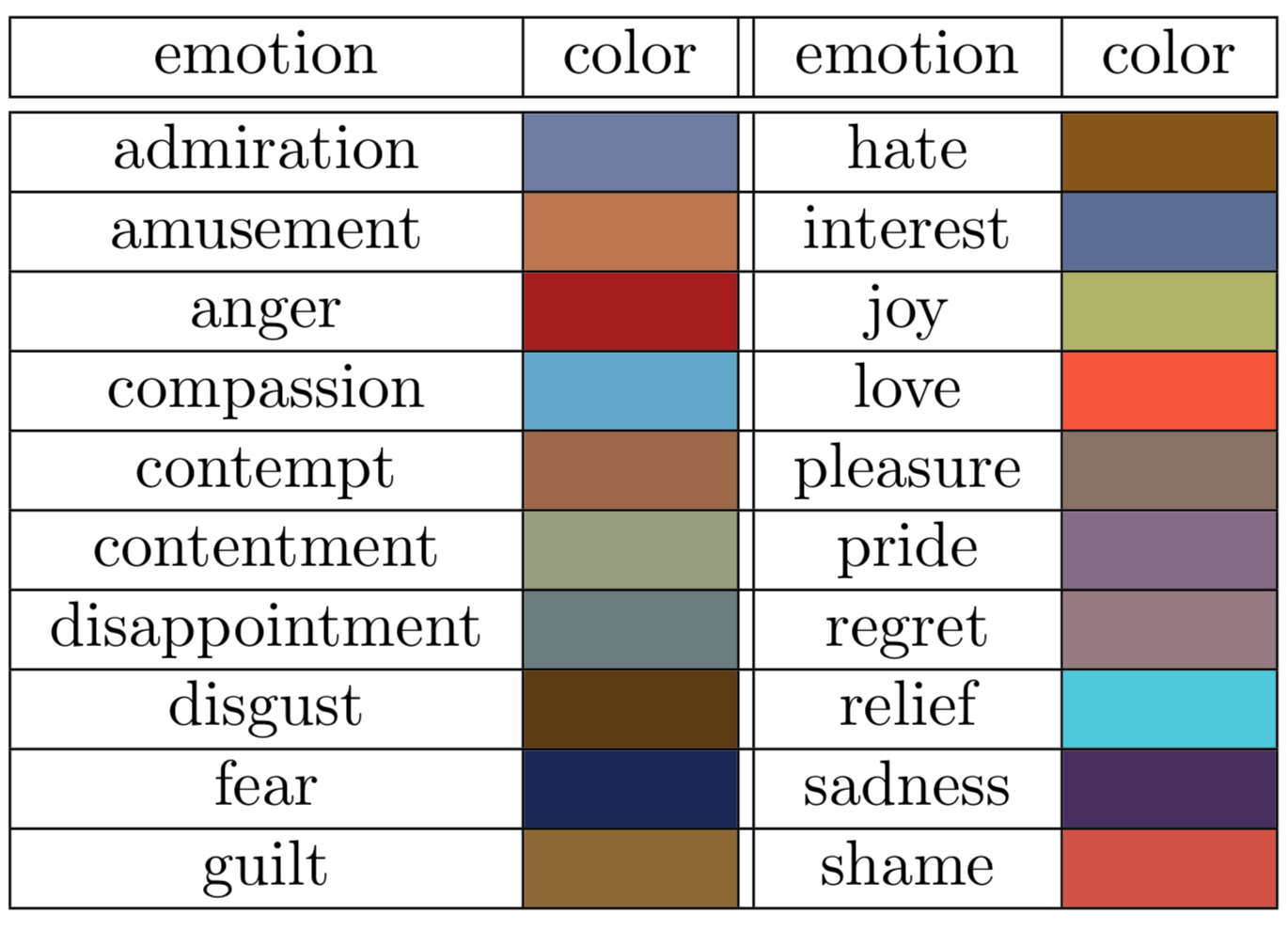}
    \end{center}
    \caption{Single emotion associations predicted through our model (Age = 20),  where shades of colours are obtained through the standard Decimal Code (R,G,B) obtained through our algorithm associated to each emotion. }
    \label{fig:table}
    \end{figure}

Through our study, we can classify emotions into two distinguished sets: those that have a distinct association to a specific colour, and those that do not.  Indeed,  emotions such as anger, disgust, fear, love, pride, relief, and sadness have very clear and distinct associations to a specific color. However, other emotions such as amusement, compassion, joy, and shame may be associated with multiple color and therefore provide an ambiguous color, generally a shade of brown or some arbitrary mix of colors.           Figure \ref{fig:table} displays the color associated with a single emotion at maximum intensity among the gender neutral and average age setting. (e.g. pure relief is most associated with a light blue color while pure anger is associated with a red color).

 \section{Concluding remarks}\label{last}
 In the present paper we have analyzed a dataset provided by the online survey \cite{plu1} in which 944 participants were presented with a series of 12 colors and asked to determine the association of each color with a set of 20 emotions in a scale from 0 to 5, via a platform depicted in Figure \ref{fig:disc}. We used a machine learning method to quantify the strength of color-emotions associations as well as their variation across age and gender. 
    To quantify the strength of color-emotion associations, we employed a neural network to classify colors based the participants 20 emotion association values as well as their age and gender. The network consists 22 input nodes and 12 output nodes (for each color), and two hidden layers of 10 nodes, as depicted in Figure \ref{fig:net}.
The results of the classification based on a cross-validation of 10000, and  on a training set of 2000, are summarized by the confusion matrix in Figure \ref{fig:conf2}. 

In particular, we found that  black, brown, pink, and red are classified correctly with high accuracy, indicating a strong association with specific emotions, while  green and purple are rarely predicted and classified accurately, indicating a weak association with emotion (see Figure \ref{assoc}). 
    In order to improve accuracy,  we combined similar colors and formed 6 color groups to reduce confusion among similar colors and ran an identical neural network to classify each group based on their corresponding emotion associations, increasing the classification accuracy to $52 \%$ (see   Figure \ref{fig:conf1red}). In particular, we find that:
   \begin{itemize}
   \item[(I)] brown is typically confused with the other darker colors, and;
\item[(II)]  both white and the orange-yellow group are notably confused with the blue-green-turquoise group.
\end{itemize}
We expect this to be likely due to shared associations with positive/negative emotions. 
    We also investigated a neural-network based regression to associate emotion and age/gender parameters colors on a continuous RGB spectrum. We used a neural network with the same 22 input nodes, but with  3 output nodes representing the red, green, and blue values. The network also consisted of 2 hidden layers of 20 and 10 nodes, and was trained on 2000 examples from the survey. We implemented an interface to display the output of the regression based on the 22 input variables, as shown in   Figure \ref{fig:RGB}. 
    
    Using the regression, we determined precise colors in  Decimal Code (R,G,B) associated with single emotions and its variation due to age and gender. The colors associated with single emotions (with age set to 20 and gender neutral) are summarized in Figure \ref{fig:table}. Most colors which were associated with multiple (conflicting) colors like amusement and shame produced ambiguous colors, generally a shade of brown or an arbitrary mix of colors. In Figure \ref{fig:table2}, we displayed the effects of gender on single emotion associations. Through  the regression, we show that:
    
   \begin{itemize}
   \item[(III)] in most cases females associated the same emotion with lighter colors, with exceptions including shame and disappointment. 
   \end{itemize}
   In particular,  Figure \ref{fig:table3} displaces the effects of age on single emotion associations.  The different perception of colours and emotional associations have long been considered (e.g., see \cite{hemphill1996note,manav2007color,boyatzis1994children}). In particular, from experimental data it is seen that  bright colors have  mainly positive emotional associations, and dark colors have mainly negative emotional associations when  not taking gender into account, but women responded more positively than men to bright colors, and they also respond more negatively to dark colors \cite{hemphill1996note}. From our mathematical study as seen in item (III) above, we see that females indeed choose brighter colours than men for most positive emotions, and choose darker colors than men for the most negative emotions, which is in agreement with the experimental results. However, there are still some negative emotions for which  women tend to choose lighter colours than men, and thus it would be very interesting to peruse an experimental study of a detailed colour association to a large range of negative emotions across gender.

   When considering age differences,  we found that  older people associate most emotions with lighter shades of the same color, with relief being an exception. It is interesting to highlight the different effects of age on colour association:
    \begin{itemize}
    \item[(IV)] Some emotions such as contempt, guilt, joy, sadness, and shame were associated with similar shades across age. 
    \item[(V)] On the other hand, some emotions like admiration and pleasure shift to a more brown color as age increases, indicating ambiguity or a lack of data (extrapolation). 
    \end{itemize}

In the last decades, many studies have found that older people   are less able to identify facial expressions  than young people \cite{phillips2002age,ebner2009young}. Moreover, since it has been hypothesised that the emotional association of colours could be related to their association with facial expressions \cite{zentner2001preferences}, one can expect the association of colour to get tinted with age, leading to further ambiguity, as we have found  in our study both through items (I) and (V) above. It would be indeed very interesting to carry out experimental research on how ageing   leads to a higher probability of brown being associated with emotions. 
    \smallskip
    
     Finally, considering the findings in terms of cross-age and gender colour association, one may consider their implications within marketing strategies of different kinds. An example of this is, for instance, when marketing perfumes for different ages and genders. In such cases, one can use the colours people tend to associate with their smells for the packaging \cite{schifferstein2004visualising}, and we propose that further considering their association with emotions could lead to 
  more targeted audiences. 
         \smallskip

    When considering single color association,  certain emotions strongly associate with multiple conflicting colors, leading to an ambiguous mix colors (often a brown) outputted by the regression. 
    This ambiguity can also explain some of the variation found in the age and gender spectrum analysis. For example, both shame and admiration have relatively weak associations across all colors, which could explain some of the variation shown in the gender comparison. It should also be noted that in  the RGB regression, different instances of the training lead to slightly different results in the colors produced. To reduce variation, we train using a high number of iterations (1000) and increase the standard regularization parameter $\lambda$ (incorporated to avoid the risk of risk of overfitting) to achieve the most consistent results. 
    
    \bigskip

 \noindent {\bf Acknowledgments.} The leading authors (*) VR and LPS   are thankful to MIT
PRIMES-USA for the opportunity to conduct this research together, and in particular Tanya Khovanova for
her continued support and James Unwin for insightful comments on a draft of the manuscript. Moreover, they greatly acknowledge the work of their co-authors to produce \cite{plu1}.\\
\pagebreak

  \noindent {\bf  Contributions of authors.} Vishaal Ram and Laura Schaposnik carried out the research and preparation of the present manuscript.  The remaining authors were responsible for the data source. In particular, Nikos Konstantinou was responsible for Greek Cypriot data of \cite{plu1}, Eliz Volkan  was responsible for Turkish Cypriot data of \cite{plu1},  Marietta Papadatou-Pastou was responsible for Greek data of \cite{plu1}, Banu Manav was responsible for Turkish data of \cite{plu1} of \cite{plu1}, and  
Christine Mohr  and Domicele Jonauskaite were esponsible for the overall data conceptualisation of \cite{plu1},
 coordination of the translations, data acquisition, and research dissemination.\\
 
  \noindent {\bf Funding.} The work of Laura Schaposnik is partially supported through the NSF grants  CAREER DMS 1749013.  The Swiss National Science Foundation supported the work of Domicele Jonauskaite with the Doc.CH fellowship grant (P0LAP1$\_$175055) and Christine Mohr with a project grant (100014$\_$182138) \\
 
 \noindent {\bf Affiliations.}\\
  (a) Milton High School, Milton, GA 30004, USA. \\
  (b)  University of Illinois, Chicago, IL 60607, USA. \\
  (c) Department of Rehabilitation Sciences
Faculty of Health Sciences
Cyprus University of Technology.\\
  (d) Cyprus International University, Nicosia, Cyprus.\\
  (e) National and Kapodistrian University of Athens, Athens, Greece.\\
  (f) Kadir Has University, Faculty of Art and Design, Department of Interior Architecture and Environmental Design. \\
  (g) Institute of Psychology, University of Lausanne, Switzerland.\\
  
\bibliography{PRIMES2020}{}

%
%
%
%



\end{document}